\documentclass[reprint,amsmath,amssymb,pra]{revtex4-1}
\usepackage{soul}
\usepackage{graphicx}
\usepackage{dcolumn}
\usepackage{bm}
\begin{document}

\title{Bragg-induced power oscillations in $\mathcal{PT}$ symmetric periodic photonic structures}

\author{P. A. Brand\~ao}
\email{paulocabf@gmail.com}
\author{S. B. Cavalcanti}%
\email{sbessa@gmail.com}
\affiliation{Universidade Federal de Alagoas, Cidade Universit\'{a}ria, Macei\'{o}-AL, 57072-970, Brazil}

\date{\today}

\begin{abstract}
We study Bragg-induced power oscillations in Fourier space between a pair of optical resonant transverse modes propagating through a periodic $\mathcal{PT}$ symmetric lattice, represented by a refractive index that includes gain and loss in a balanced way. Our results imply that the $\mathcal{PT}$-symmetric system shows exceptionally rich phenomena absent in its Hermitian counterpart. It is demonstrated that the resonant modes exhibit unique characteristics such as Bragg power oscillations controlled via the $\mathcal{PT}$ symmetry, severe asymmetry in mode dynamics, and trapped enhanced transmission. We have also performed numerical simulations in (1+1) and (2+1) dimensions of propagating Gaussian beams to compare with analytical calculations developed under a two-waves model.
\end{abstract}

\maketitle


\section{\label{intro}Introduction}

More than a decade ago, a very interesting idea has been put forward by Bender and co-workers on the existence of non-Hermitian Hamiltonians exhibiting a real-valued energy spectrum below a symmetry breaking point at which, the spectrum undergoes a phase transition to the complex plane \cite{bender,bender01,bender02,bender03,bender04}. These non-Hermitian Hamiltonians are very useful for modelling optical systems with a controlled balance between amplification and attenuation, meaning that the refractive index obeys the property $n(x) = n(-x)^{*}$. Therefore the $\mathcal{PT}$ symmetry concept was soon extended to optical systems by identifying the real part of the potential with the refractive index of the lattice and the imaginary one with gain/loss. Complex potentials have received then, in recent years, an enormous attention considering the development of these new artificial structures where one may tailor the refractive index according to one's need. Theoretical work based on the paraxial approximation have reported results on the propagation of light beams transversely to an optical lattice, that indicate that the input beam power is not conserved along propagation \cite{Berry}. Experimental work based on $\mathcal{PT}$ Hamiltonians have been suggested in optical media with a complex refractive index \cite{muga,makris,Moiseyev,Bendix} and, $\mathcal{PT}$ symmetry breaking has been firstly reported in experiments based on a passive optical coupler \cite{Guo}. Bloch oscillations have been proposed \cite{Longhi2009} and demonstrated in $\mathcal{PT}$ symmetric photonic lattices \cite{Bender15}, exhibiting rich phenomena not present in conservative systems. Also, experimental reports on light transport in $\mathcal{PT}$ symmetric temporal-lattices have demonstrated that such symmetric periodic structures might act as unidirectional invisible media near a spectral singularity \cite{invisibility}. Spectral singularities are resonant energy eigenvalues that represent states characterized by infinite reflection and transmission coefficients. Spontaneous $\mathcal{PT}$ symmetry breaking and non-reciprocal power oscillations have been observed in a study on the behavior of a $\mathcal{PT}$ optical coupled system \cite{Ruter}.  Overall, one may conclude that the applications of $\mathcal{PT}$ symmetry to optical systems should provide new ways to control light propagation, unraveling a world of remarkable effects with tremendous impact on optical materials, devices and networks \cite{Nature}.

Motivated by these exciting reports, in this work we wish to investigate the role of $\mathcal{PT}$ symmetry on optical power oscillations, due to the propagation of wide optical beams in complex periodic lattices, whereby the coupling between the beam and the periodic structure results into Bragg-induced power oscillations between transverse modes related by the Bragg resonance condition \cite{valery,nosso}. More specifically, we wish to investigate the consequences of extending the model presented in \cite{valery} to the complex domain. It is well-known that by doing this, one opens up the way to strange and fascinating properties, particularly on the verge of spectral singularities. In this case, a close examination on Bragg scattering processes has demonstrated that spectral singularities are associated with the secular growth of plane waves that satisfy the Bragg condition and also that they occur at the $\mathcal{PT}$ symmetry breaking point. Furthermore, it has been reported that, in contrast with wide beams, a wave packet with a broad momentum distribution leads to a saturation of the secular growth of scattered waves \cite{longhispectral,eva}. In the following we choose a $\mathcal{PT}$ potential periodic function to investigate the influence of $\mathcal{PT}$ symmetry on the behavior of Bragg-induced power oscillations between a pair of Bragg-resonant modes excited by wide optical beams. Based on the two-waves model complemented by numerical simulations of the wave equation by using a Gaussian beam input, we investigate the Bragg-resonant mode behavior below and above the phase transition point, and we obtain all the peculiar universal features of non-Hermitian Hamiltonians, such as nonreciprocal behavior of the Bragg modes, power oscillations, and secular growth of plane waves. Furthermore, we find new features on the spectral behavior such as mode trapping and asymmetric mode power transfer. 

\section{Two-waves model}

It is well known that a plane wave input in general may excite many resonant lattice modes. Here, we suppose that the input power mode is Bragg-resonant with the lattice so that the incident plane wave excites mainly the two-modes at the edges of the Brillouin zone and ignore all other resonant modes \cite{Yariv}. More specifically, we consider a one dimensional periodic $\mathcal{PT}$ symmetric potential $V(x)$ function through which a wide beam $\psi(x,z)$ propagates. The propagation dynamics is then described by the paraxial wave equation
\begin{equation}
\label{schr}
i\frac{\partial\psi}{\partial z} + \frac{\partial^{2}\psi}{\partial x^{2}} - V(x)\psi = 0,
\end{equation}
in arbitrary units. Let us define a potential function of the form $V(x) = \alpha[\cos^{2}x + i\beta\sin(2x)]$ with $\alpha,\beta$ representing real parameters (with $\beta$ positive). It is easy to verify that the so-defined potential satisfies $V(-x)^{*} = V(x)$, and there is a gain-loss parameter $\beta$ that controls the degree of Hermiticity. This parameter defines a spontaneous symmetry breaking point $\beta_{c} = 1/2$, above which the spectrum undergoes a phase transition from the real to the complex plane. More specifically, the spectrum is real for $\beta < \beta_{c}$ and becomes partially or completely complex for $\beta > \beta_{c}$ \cite{makris}. We solve the wave equation by writing the field as,
\begin{equation}
\label{eq2}
\psi(x,z) = \sum_{n=\pm 1,\pm 2 ...} \psi_{n}(z)\exp(ink_{b}x),
\end{equation}
where $k_{b} = 1$ is the resonant transverse wavevector and the index $n$ indicates the $n$th mode that is Bragg resonant with the lattice. Substituting (\ref{eq2}) into (\ref{schr}), one arrives at the following set of coupled differential equations,
\begin{equation}
\label{sys1}
i\frac{d\psi_{n}}{dz} = a_{n}\psi_{n} + b\psi_{n-2} + c\psi_{n+2}
\end{equation}
with 
\begin{equation}
a_{n} = \left(n^{2}+\frac{\alpha}{2}\right)
\end{equation}
and
\begin{equation}
\label{eqb}
b = \frac{\alpha}{2}(\beta_{c}+\beta),
\end{equation} 
\begin{equation}
\label{eqc}
c = \frac{\alpha}{2}(\beta_{c}-\beta).
\end{equation} 
Here we study the particular case where $n = \pm1$ only. This is the essence of the two-waves model and it was shown to give reliable results when the incident angle is coupled to the edge of the first Brillouin zone \cite{valery}. Its validity will be further discussed in Section V. Eqs. (\ref{sys1}) are first order in $z$ so that the initial values $\psi_{-1}(0)$ and $\psi_{1}(0)$ must be specified for a unique solution. To acquire a complete picture of the dynamics we now consider the following second order linear differential equation for $\psi_{-1}$,
\begin{equation}
\label{eq5}
\frac{d^{2}\psi_{-1}}{dz^{2}} + 2ia\frac{d\psi_{-1}}{dz} + (cb - a^{2})\psi_{-1}=0 ,
\end{equation}
where $a = a_{-1} = a_{1}$. Eq. \eqref{eq5} is derived from the system of equations \eqref{sys1} for $n = \pm 1$ and considering $\psi_{-3} = \psi_{3} = 0$. The equation satisfied by $\psi_{1}(z)$ is obtained from \eqref{eq5} by performing the operation $\psi_{-1}\rightarrow\psi_{1}$. The general solutions for the modal amplitudes $\psi_{-1}(z)$ and $\psi_{1}(z)$ are given by:
\begin{multline}
\label{eq9}
\left\{\substack{\psi_{-1}(z) \\ \psi_{1}(z)}\right\} = \frac{1}{2}\left[ \left\{\substack{\psi_{-1}(0) \\ \psi_{1}(0)}\right\} -i\left\{\substack{\gamma\psi_{1}(0) \\ \gamma'\psi_{-1}(0)}\right\}\right] \\
\times \exp\left[ -\left( \frac{\alpha}{2} + 1 \right)iz + \frac{|\alpha|}{2}kz\right] \\
+ \frac{1}{2}\left[ \left\{\substack{\psi_{-1}(0) \\ \psi_{1}(0)}\right\} +i\left\{\substack{\gamma\psi_{1}(0) \\ \gamma'\psi_{-1}(0)}\right\}\right] \\
\times \exp\left[-\left( \frac{\alpha}{2} + 1 \right)iz - \frac{|\alpha|}{2}kz\right],
\end{multline}
where 
\begin{equation}
\label{eq11}
\gamma = \text{sgn}(\alpha) \frac{(\beta-\beta_{c})}{k},
\end{equation}
\begin{equation}
\label{eq111}
\gamma' = \text{sgn}(\alpha) \frac{(\beta+\beta_{c})}{k},
\end{equation}
$k = \sqrt{\beta^{2} - \beta_{c}^{2}}$ and $\text{sgn}(\alpha)$ is $+1$ or $-1$ depending on the sign of $\alpha$. The initial conditions $\{\psi_{-1}(0),\psi_{1}(0) \}$ together with relations \eqref{eq9}, \eqref{eq11} and \eqref{eq111} determine the system's evolution. Before we discuss some more involved peculiarities, note the apparent asymmetry in the evolution of the two modal functions that is solely due to the $\gamma$ and $\gamma'$ terms. This parameter may be either real or complex depending on the system being below ($\beta < \beta_{c} $) or above ($\beta > \beta_{c}$) the critical phase point. To fully grasp the system's dynamics, in the following we separate the discussion in these two cases.

Before we proceed, it is useful to define the population inversion function $W(z)$ which characterizes the spectral energy exchange between the two resonant modes. We define $W(z) = |\psi_{1}(z)|^{2} - |\psi_{-1}(z)|^{2}$ representing the difference in power spectra of the two modes. A $\mathcal{PT}$ symmetric system, in general, does not satisfy symmetric initial conditions and thus, one must distinguish two cases depending on the values of $\psi_{-1}(0)$ and $\psi_{1}(0)$. Let us begin by choosing $\{\psi_{-1}(0),\psi_{1}(0)\} = \{1,0\}$, so that, 
\begin{multline}
\label{eq12}
  W_{-1} = \frac{1}{4}\left[ \left( |\gamma'|^{2}-1\right)e^{-(k+k^{*})z} \right.\\
    \left. - \left( 1+|\gamma'|^{2}\right)e^{-(k-k^{*})z} \right. \\
    \left. - \left( 1+|\gamma'|^{2}\right)e^{(k-k^{*})z} \right. \\
    \left. + \left( |\gamma'|^{2}-1\right)e^{(k+k^{*})z}\right].
\end{multline}
On the other hand, by choosing $\{\psi_{-1}(0),\psi_{1}(0)\} = \{0,1\}$, the population inversion becomes 
\begin{multline}
\label{eq13}
  W_{1} = \frac{1}{4}\left[ \left( 1-|\gamma|^{2}\right)e^{-(k+k^{*})z} \right.\\
    \left. + \left( |\gamma|^{2}+1\right)e^{-(k-k^{*})z} \right. \\
    \left. + \left( |\gamma|^{2}+1\right)e^{(k-k^{*})z} \right. \\
    \left. + \left( 1-|\gamma|^{2}\right)e^{(k+k^{*})z}\right].
\end{multline}
These expressions form the basis for the discussion of population inversion in systems with $\mathcal{PT}$ symmetry within the two-waves approach and will be used extensively in the next sections.

\section{Population inversion below the phase transition point}

\begin{figure}[ht]
\includegraphics[width=9cm]{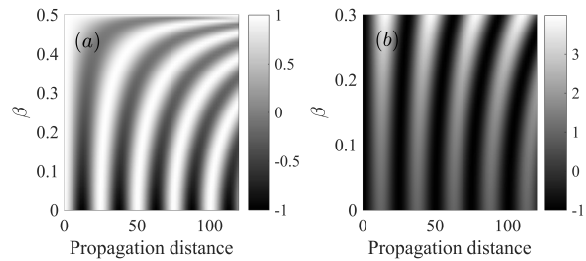}
\centering
\caption{(a) Population inversion $W_{1}$ in a $\mathcal{PT}$ symmetric lattice as a function of the order parameter $\beta$ and propagation distance, for $\beta < \beta_{c}$ illustrating power oscillations. (b) Same for $W_{-1}$ with $\beta$ chosen to be in the range $[0,0.3]$ to make the color scale easier for visualization. Notice how the color scale goes beyond the value of 1. For this picture we have assumed $\alpha = 0.5$.}
\label{fig1}
\end{figure}

\begin{figure}[ht]
\includegraphics[width=9cm]{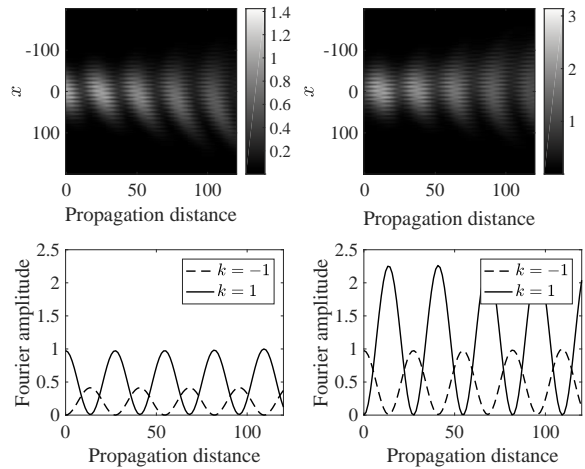}
\centering
\caption{Gaussian beam propagation below the phase transition point ($\beta < \beta_{c}$) in both real (upper panels) and Fourier spaces (lower panels) illustrating power oscillations with $\beta = 0.2$ and $\alpha = 0.5$. Left column: $\{\psi_{-1}(0),\psi_{1}(0) \} = \{0,1\}$. Right column: $\{\psi_{-1}(0),\psi_{1}(0) \} = \{1,0\}$. Excited modes outside the first Brillouin zone are indistinguishable from the horizontal zero axis line.}
\label{fig2}
\end{figure}

In the case $0 \leq \beta \leq \beta_{c}$ one may define $k = i\kappa$ with $\kappa = \sqrt{\beta_{c}^{2}-\beta^{2}} \geq 0$ a real non-negative constant. Let us first consider a real lattice for which $\beta = 0$. Then, $|\gamma|^{2} = |\gamma'|^{-2} = 1$, $k = i\beta_{c}$ and Eqs. (\ref{eq12}) and (\ref{eq13}) give $W_{-1} = -\cos (|\alpha|z/2)$ and $W_{1} = \cos(|\alpha|z/2) = -W_{-1}$, which are symmetric so that the system executes full power spectra oscillations with a complete periodic transfer of energy between the two modes at every $z_{p} = 4\pi/|\alpha|$ \cite{valery}. Things become very different for a complex lattice. To illustrate how the population inversion evolves in this scenario, we choose  $0 \leq \beta \leq \beta_{c}$, so that Eqs. (\ref{eq12}) and (\ref{eq13}) become
\begin{equation}
\label{eq15}
W_{1} =\cos^{2}\left(\frac{|\alpha|\kappa z}{2}\right)\\
- \frac{(\beta_{c}
- \beta)}{(\beta_{c}+\beta)}\sin^{2}\left(\frac{|\alpha|\kappa z}{2}\right),
\end{equation}
\begin{equation}
\label{eq16}
W_{-1} = -\cos^{2}\left(\frac{|\alpha|\kappa z}{2}\right)\\
+ \frac{(\beta_{c}
+ \beta)}{(\beta_{c}-\beta)}\sin^{2}\left(\frac{|\alpha|\kappa z}{2}\right),
\end{equation}
which clearly highlights the asymmetry of the system due to the factors $\gamma$ and $\gamma'$. The two-parameter families of functions $W_{-1}(\beta,z)$ and $W_{1}(\beta,z)$ may be visualized in color plots to illustrate
power spectra oscillations when the lattice is below the phase transition point.  Let us turn to Figure \ref{fig1}, where $W_{1}$ and $W_{-1}$ are plotted respectively in Figures \ref{fig1}(a) and \ref{fig1}(b) from which many interesting conclusions are drawn. Figure \ref{fig1}(a) shows that if $\beta \approx 0$, one retrieves the conservative power oscillation as a function of propagation distance and here described by $W_{1} \approx \cos( |\alpha|z/2)$, with colours ranging from $+1$ (white) to $-1$ (black) for all $z$. This means that power oscillations are represented here by the successive regions of white and black, in the sense that power is more present in mode $\psi_{1}$ in white regions and is more present in mode $\psi_{-1}$ in dark ones. Therefore in Figure 1 white regions means more power within mode $\psi_{1}$ and dark regions means more power within mode $\psi_{-1}$. Gray regions means that the power is shared between the two modes in Figure \ref{fig1}(a). On the other hand, Figure \ref{fig1}(b) shows that if the initial power spectrum is in the mode $\psi_{-1}$, oscillations are retrieved although the spectra amplitudes are much larger in mode $\psi_{1}$ than the previous case, indicating that power transfer is much more efficient to mode $\psi_{1}$ than to mode $\psi_{-1}$ (see also Figure \ref{fig2}).  As $\beta$ approaches the critical value $\beta_{c}$, there is a distortion of the vertical lines in Figure \ref{fig1}, that is mainly due to the term $\sqrt{\beta_{c}^{2}-\beta^{2}}$ inside the trigonometric functions in Eqs. (\ref{eq15}) and (\ref{eq16}). To see more clearly what is actually happening at the critical point, let us substitute $\beta = \beta_{c}$ directly in Eq. (\ref{eq15}) to obtain
\begin{equation}
\label{eq17}
W_{1}(z) = 1.
\end{equation}
Exactly at the transition point, when the input power is located initially in mode $\psi_{1}$, it will remain trapped in this mode forever independent of the propagation distance. There is a perfect balance of energy exchange between lattice and wavefield such that a stable intensity evolution is achieved. When $\beta$ is smaller than $\beta_{c}$, power oscillations are modulated, through the $\mathcal{PT}$ symmetry of the lattice, with an increase in the oscillation period. This behavior is evident by a close inspection of Figure \ref{fig1} and also by Eq. (\ref{eq15}) where $\beta$ appears inside the sine (cosine) and thus controls the oscillation period together with the value of $|\alpha|$.
On the other hand, when the initial state of the system is defined by $\{\psi_{-1}(0),\psi_{1}(0) \} = \{1,0\}$, the parameter $\gamma'$ suggests some peculiar behavior and, by taking the limit $\beta\rightarrow\beta_{c}$ directly, as was done earlier, we obtain from \eqref{eq16}: 
\begin{equation}
W_{-1} = -1 + \frac{|\alpha|^{2}}{4}z^{2},
\end{equation}
where we used the fact that $\lim_{p\rightarrow 0}\sin^{2}(\sqrt{p}y)/p = y^{2}$. The population difference exhibits secular growth during propagation when $\beta = \beta_{c}$ exactly. This means that the lattice is transferring energy preferably to mode $\psi_{1}(z)$, because $W_{-1}$ will certainly become positive. To better visualize this, one may write the total field $\eqref{eq2}$ at $\beta = \beta_{c}$ [with $\{\psi_{-1}(0),\psi_{1}(0)\} = \{1,0\}$]: 
\begin{equation}
\label{eq19}
\psi(x,z) = e^{-i(1+\alpha/2)z}\left[ \text{sgn}(\alpha)
\frac{\alpha z}{2i}e^{ix} + e^{-ix}\right],
\end{equation}
where it is easy to see that, indeed, the first amplitude factor, which is responsible for the mode $\psi_{1}(z)$, grows linearly with the propagation distance, while the mode at $k = -1$ remains constant. 

\begin{figure}
\includegraphics[width=9cm]{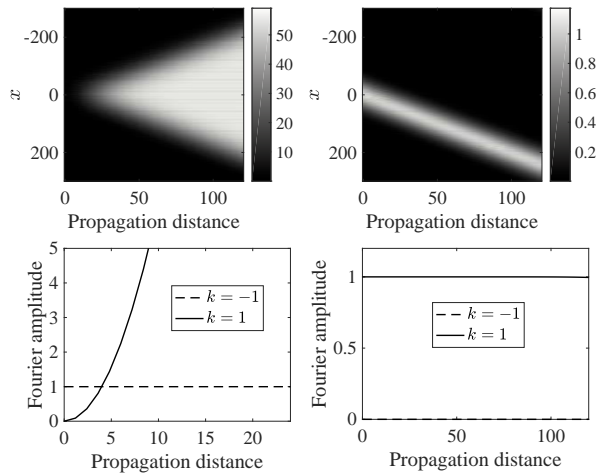}
\centering
\caption{Wide Gaussian beam propagation at the critical phase point ($\beta = \beta_{c}$) in both real 
(upper panels) and Fourier (lower panels) spaces, with initial width $X_{0} = 15p$ ($p$ is the lattice period) incident on a $\mathcal{PT}$ symmetric lattice at the critical point. Left column: $\{\psi_{-1}(0),\psi_{1}(0) \} = \{1,0\}$. Right column: $\{\psi_{-1}(0),\psi_{1}(0) \} = \{0,1\}$. Excited modes outside the first Brillouin zone are indistinguishable from the zero horizontal axis line. }
\label{fig3}
\end{figure}

\begin{figure}[ht]
\includegraphics[width=7cm]{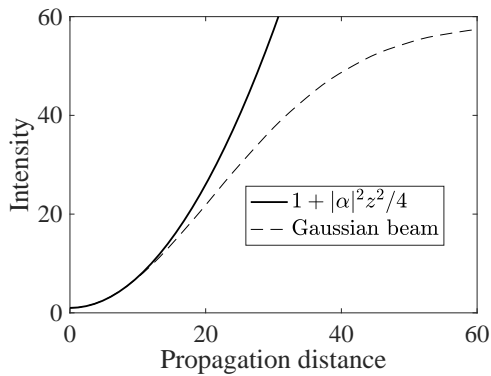}
\centering
\caption{Intensity at $x = 0$ as a function of the propagation distance with the lattice at the critical point. The dashed line represents the intensity obtained from the simulation with a Gaussian beam. Continuous line is the analytic result from the two-waves model [Eq. \eqref{eq19}]. For this simulation we have assumed $\alpha = 0.5$.}
\label{fig4}
\end{figure}

Next, we compare numerical results based on the propagation of a wide Gaussian beam input with spatial spectral power centered at the edge of the Brillouin zone, with the analytic ones obtained from the two-waves model. To this end we solve Eq. (\ref{schr}) using a split-step Fourier method with a Strang splitting, with the input 
\begin{equation}
\psi(x,0) = \exp\left[-\frac{1}{2}\left(\frac{x}{X_{0}}\right)^{2}\right]\exp(\pm ix),
\end{equation}
$\alpha = 0.5$ and initial width $X_{0} = 15p$, where $p$ is the lattice period. Power oscillations are illustrated in Figure \ref{fig2} where the propagation of a Gaussian beam, in both real (upper panels) and Fourier (lower panels) spaces is shown, below the transition point, whether the initial condition is $\{\psi_{-1}(0),\psi_{1}(0) \} = \{0,1\}$ and $\{\psi_{-1}(0),\psi_{1}(0) \} = \{1,0\}$ presented in left and right panels respectively. The asymmetry regarding the transfer of power spectra between the two modes is clearly evident by inspecting the Fourier amplitudes. Note the striking contrast between the dynamics in these two cases. The initial Fourier spectrum is centered at $k_{i} = -1$ (left panels) or at $k_{i} = 1$ (right panels). Excited modes outside the first Brillouin zone, at $k = \pm 3$ and beyond, are indistinguishable from the horizontal zero axis line (not shown). Figure \ref{fig3} shows the numerical result of the Gaussian beam propagation in both real (upper panels) and reciprocal (lower panels) spaces, with $\beta = \beta_{c}$. When $\{\psi_{-1}(0),\psi_{1}(0) \} = \{0,1\}$ (right panels) one may clearly see that the initial Gaussian beam propagates along a definite angle with a steady intensity. This result is in good agreement with the analytic result predicted by Eq. (\ref{eq17}). On the other hand, the initial condition $\{\psi_{-1}(0),\psi_{1}(0) \} = \{1,0\}$ (left panels) indicates that in this case, the lattice is giving energy to the field and, as a result, the intensity increases indefinitely during propagation. Notice the dominance of the $\psi_{1}(z)$ mode, no matter where the input is, energy always flows higher to this mode.  Figure \ref{fig4} shows the intensity as function of the propagation distance at $x=0$. Due to the previous analysis, one expects that the intensity evolves according with $|\psi(0,z)|^{2} = 1 + (|\alpha|z)^{2}/4$ and we compare this evolution, in long propagation distances, predicted by the two-waves model (solid line) with the evolution of a Gaussian beam (dashed line). We find that in this case the contribution of other modes apart from the two considered here, indicates a tendency to the spectral broadened induced saturation effect that is known to neutralize secular growth reported in  \cite{longhispectral} (dashed line in Figure \ref{fig4}). Nevertheless, the initial overall behavior before the saturation effect takes place, and the role of the $\mathcal{PT}$ symmetric lattice, are clearly captured by the two-waves model.

\begin{figure}[ht]
\includegraphics[width=6.5cm]{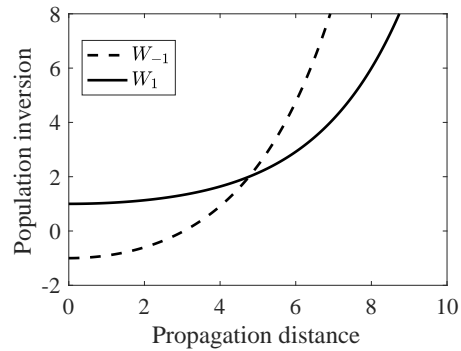}
\centering 
\caption{Population inversion for a system above the phase transition point as function of the propagation distance $z$ with with $\beta = 1$ and $\alpha = 0.5$. Dashed line: $W_{-1}$. Solid line: $W_{1}$. Clearly, the lattice energy flows towards mode $\psi_{1}(z)$ is dominant.}
\label{fig5}
\end{figure}

\begin{figure}[ht]
\includegraphics[width=7cm]{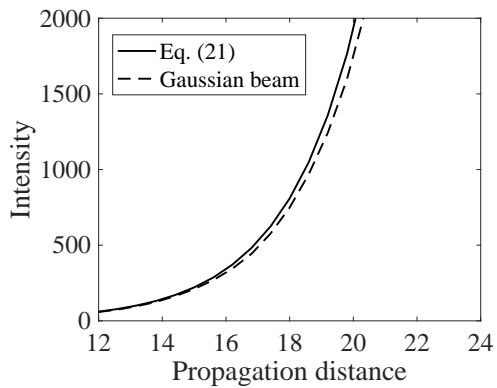}
\centering
\caption{Field intensity evolution at $x=0$ of a wavefield during propagation above the phase transition point with  the same parameters as Figure 5. The dashed line represents the intensity from the numerical simulation with a Gaussian beam. Solid line is obtained from the two-waves approach.}
\label{fig6}
\end{figure}

\section{Population inversion above the phase transition point}

When $\beta \geq \beta_{c}$ the system has undergone a phase transition meaning that eigenvalues of the Hamiltonian operator in Eq. (\ref{schr}) have become complex or partial complex. This has a tremendous consequence on the population inversion described by Eqs. (\ref{eq12}) and (\ref{eq13}), which now reads, 
\begin{equation}
\label{eq20}
W_{-1} = -\cosh^{2}\left( \frac{|\alpha|kz}{2} \right) \\
+ \frac{(\beta+\beta_{c})}{(\beta-\beta_{c})}\sinh^{2}\left( \frac{|\alpha|kz}{2} \right),
\end{equation}
\begin{equation}
\label{eq21}
W_{1} = \cosh^{2}\left( \frac{|\alpha|kz}{2} \right) \\
- \frac{(\beta-\beta_{c})}{(\beta+\beta_{c})}\sinh^{2}\left( \frac{|\alpha|kz}{2} \right),
\end{equation}
clearly displaying uncontrolled diverging behaviors.  After a close inspection of Eqs. (\ref{eq20}) and (\ref{eq21}) one concludes that the energy transfer process is not symmetric with respect to the excited modes, in the sense that the energy flow is always directed towards $\psi_{1}(z)$, no matter the initial condition one chooses. This asymmetric mode evolution is typical of $\mathcal{PT}$ symmetric Hamiltonians. In Figure \ref{fig5} we illustrate this statement by plotting $W_{-1}$ (dashed line) and $W_{1}$ (solid line) for $\beta = 1$ and $\alpha = 0.5$ as functions of the propagation distance $z$. The asymmetry in the dynamics of the pair is evident in view of a privileged  mode namely $\psi_{1}$ for which, power always flows, whether the initial state is populated or not. 

Let us now compare the field intensity evolution of the analytic approach with the Gaussian beam, as we did in the previous section. By  considering $\{\psi_{-1}(0),\psi_{1}(0)\} = \{0,1\}$, it is straightforward to show that:
\begin{equation}
\label{eq22}
|\psi(0,z)|^{2} = \frac{\beta}{\beta+\beta_{c}}\cosh \left(|\alpha|\sqrt{\beta^{2}-\beta_{c}^{2}} z \right) + \frac{\beta_{c}}{\beta+\beta_{c}},
\end{equation} 
 which means that, above the phase transition point, the model predicts an exponential growth for the intensity, controlled by the hyperbolic cosine function. Figure \ref{fig6} depicts the field evolution predicted by \eqref{eq22} compared with a Gaussian initial condition. Thus, one may conclude that the lattice is giving energy to the field during propagation 
and that this energy transfer is even more pronounced when the system is above the phase transition point.

\section{The two-waves approximation}

\begin{figure}[ht]
\includegraphics[width=7cm]{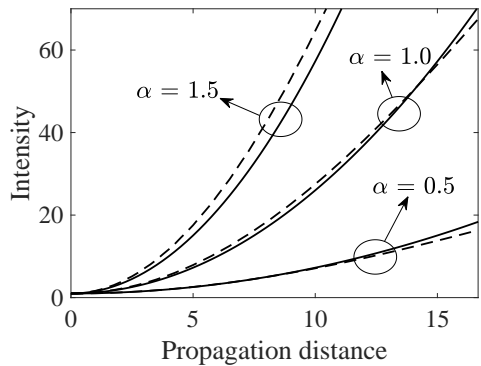}
\centering
\caption{(Dashed lines) Gaussian beam intensity evolution at $x =
0$ as a function of the propagation distance for three values of
$\alpha$. (Solid lines) Corresponding two-mode approach
$1 + (|\alpha|z/2)^2$.}
\label{fig7}
\end{figure}

\begin{figure}
\includegraphics[width=9cm]{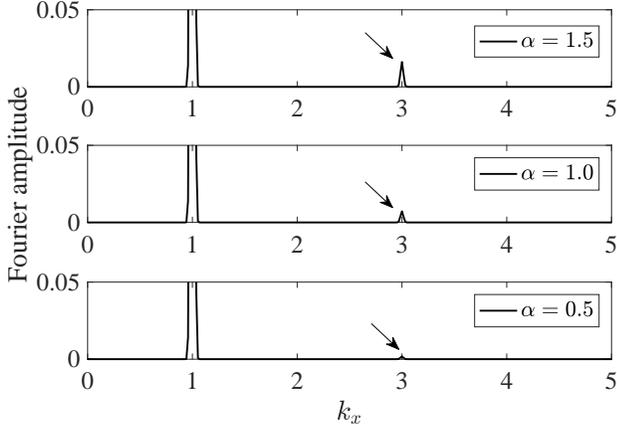}
\centering
\caption{Fourier amplitudes correspondent to the intensity depicted in Figure \ref{fig7} for the same values of $\alpha$ at $z = 15$. The plot is normalized 
such that the maximum value of the Fourier amplitude at $k_{x} = 1$ is 1.}
\label{fig8}
\end{figure}

\begin{figure}
\includegraphics[width=9cm]{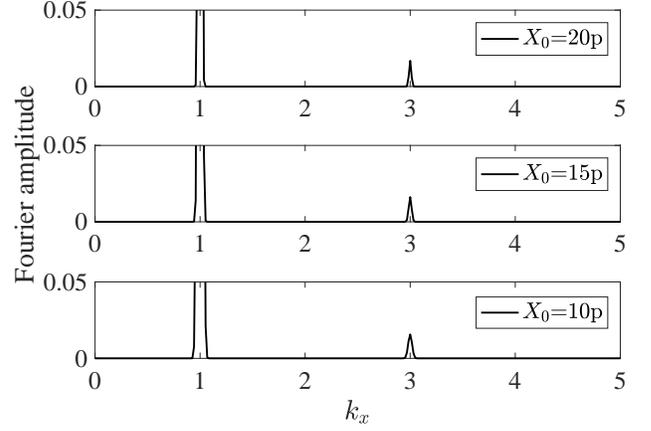}
\centering
\caption{Fourier amplitude with fixed $\alpha = 1.5$ and various input 
beam widths (part (a) of Figure \ref{fig8}) and (a) $X_{0} = 20p$, (b) $X_{0} = 15p$, (c) $X_{0} = 10p$. The plot is normalized such that the maximum value 
of the Fourier amplitude at $k_{x} = 1$ is 1.}
\label{fig9}
\end{figure}

One of the main reasons for the contrasting behavior exhibited in Figures 4 and 6 stems from the fact that the two-waves approximation implies a shallow potential function, for it determines the lattice coupling strength with other beam modes, besides the pair considered. This strength is represented here by the overall amplitude coefficient $\alpha$, that we have assumed equal to 0.5 in the previous discussed results. At the symmetry breaking point, $\beta = \beta_{c}$, the intensity at $x = 0$ is obtained from Eq. \eqref{eq19}:
\begin{equation}
|\psi(0,z)|^{2} = 1+\frac{|\alpha|z^{2}}{4}.
\end{equation}
So, the parameter $\alpha$ represents a correction to the parabolic curvature of the intensity. 
This result is illustrated in Figure \ref{fig7} where the evolution of the intensity at $x=0$ of a Gaussian input with the propagation distance is shown for three values of $\alpha$. It is clear that, as long as $|\alpha| \lesssim 1$, the analytic model reaches a better agreement with more realistic beam sources for a determined propagation distance. This is expected because the propagating beam excites few modes for $|\alpha| \ll 1$. In contrast, larger values of $\alpha$ lead to the coupling with other modes. To see the coupling effect promoted by $\alpha$, we have plotted in Figure \ref{fig8} the spectrum power 
correspondent to the intensity shown in Figure \ref{fig7}, for $z=15$, where the appearance of another mode at $k_{b} = 3$ is clear. For this graph we have normalized the maximum value of the Fourier modes so that the maximum value of the mode seen at $k_{b} = 1$ is 1. Note that increasing the absolute value of $\alpha$ leads to the enhancement in the amplitude of the additional Fourier mode and the mode at $k_{b} = 2$ is not excited at all for the lattice couples with odd integers only, $\pm1$, $\pm3$, etc, [See Eq. \eqref{sys1}]. The coupling effect is also apparent in Figure \ref{fig9}, where the Fourier amplitudes corresponding to Figure \ref{fig8}(a), i.e., 
for fixed $\alpha=1.5$ and $z=15$, and various beam  widths are depicted. Here, it should be noted that the only effect promoted by narrow input beams 
is to broaden the resonant Fourier amplitudes.  However, their amplitudes keep still as the maximum value of their Fourier amplitude is not modified. Particularly, the additional mode at $k_{b} = 3$ becomes wider but not higher, and so one might conjecture that this fact could be the reason why the analytic approach yields so reliable results. More specifically it suggests that secondary modes are less efficiently excited due to the spectral 
broadening. In this way,  one concludes that this maximum is determined solely by the value of $\alpha$ and that efficient coupling is determined fundamentally by the potential strength. This result is quite reasonable by having in mind that a finite input beam, in spite of providing additional 
modes to be excited, does not promote the coupling, which is not going to happen when the potential is too weak to induce secondary expressive scattering processes. Therefore, the two-waves model in spite of its simplicity describes quite well the overall behavior of light propagation in a complex lattice in the case of shallow potential functions and multimode optical fields. 

\section{Spectral power oscillations in (2+1) dimensions}

\begin{figure}
\includegraphics[width=9cm]{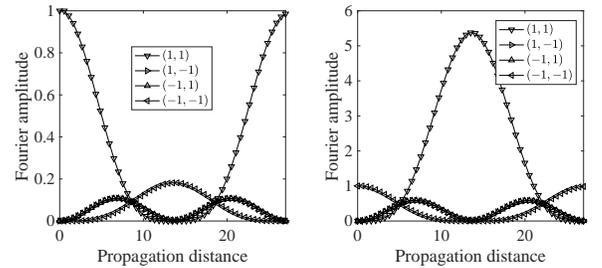}
\centering
\caption{Bragg-induced power oscillations in two dimensional beams with 
the lattice below the symmetry breaking point for $\alpha = 0.5$ and $\beta =  0.2$. Evolution of the Fourier 
amplitudes when the Fourier spectrum of the incident beam is located at (left) $(k_{bx},k_{by}) = (1,1)$ and (right) $(k_{bx},k_{by}) = (-1,-1)$. This plot is normalized such that the initial Fourier amplitude in both cases is equal to unity.}
\label{fig10}
\end{figure}

\begin{figure}
\includegraphics[width=9cm]{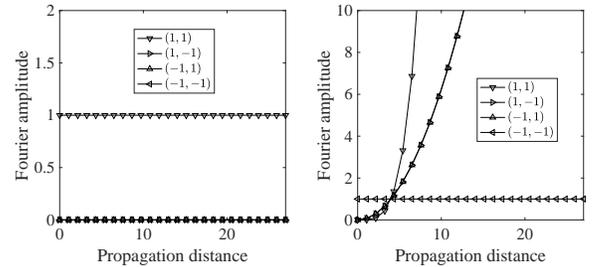}
\centering
\caption{Two dimensional mode trapping in a $\mathcal{PT}$ symmetric lattice for 
$\alpha = 0.5$ and $\beta =  0.5$. The Fourier amplitude of the incident beam is located at (left) $(k_{bx},k_{by}) = (1,1)$ and (right) $(k_{bx},k_{by}) = (-1,-1)$. This plot is normalized such that the initial Fourier amplitude in both cases is equal to unity.}
\label{fig11}
\end{figure}

Let us now find whether our analysis can still give reasonable results when extended to higher dimensions. To this end, in the following we wish to investigate the mode trapping effect at the symmetry breaking point, observed above, considering a propagating two-dimensional (2+1) optical field $\psi(x,y,z)$. We then write the two dimensional version of Eq. \eqref{schr}. 
\begin{equation}
\label{eq23}
i\frac{\partial\psi}{\partial z} + \frac{\partial^{2}\psi}{\partial x^{2}} + \frac{\partial^{2}\psi}{\partial y^{2}} - V(x,y)\psi = 0,
\end{equation}
with the $\mathcal{PT}$ symmetric potential \cite{makris}
\begin{equation}
V = \alpha\left\{ \cos^{2}x+\cos^{2}y + i\beta[\sin(2x) + \sin(2y)] \right\}.
\end{equation}
In a two-dimensional system there are four modes in the first Brillouin zone that are effective coupled so that in 2D systems, one must use the four waves model instead. By expanding the field according to Eq. \eqref{eq2},
\begin{equation}
\label{eq26}
\psi(x,y,z) = \sum_{n,m=\pm1,\pm2,...}\psi_{n,m}(z)e^{inx+imy},
\end{equation}
and inserting Eq. \eqref{eq26} into Eq. \eqref{eq23} one obtains the following coupled linear differential equations for the modes $\psi_{n,m}(z)$:
\begin{multline}
i\frac{d\psi_{n,m}}{dz} = a\psi_{n,m} + b(\psi_{n-2,m} + \psi_{n,m-2}) \\
+ c(\psi_{n+2,m} + \psi_{n,m+2}) = 0,
\end{multline}
where $a = (n^{2}+m^{2}+\alpha)$ and $b$ and $c$ are given by \eqref{eqb} and \eqref{eqc}, respectively. Although analytic solutions to this more general 
case are still possible, they are quite cumbersome and, therefore, we attempt 
to solve Eq. \eqref{eq23} numerically with a Gaussian initial condition to verify the mode trapping phenomenon at the symmetry breaking point. More specifically, we suppose
that the coupling is effective only at the corners of the first Brillouin zone,
so that the initial Gaussian beam amplitude, $\psi(x,y,0)$, may be written as
\begin{equation}
\label{eq25}
\psi(x,y,0)= \exp\left[ -\frac{(x^{2}+y^{2})}{2X_{0}^{2}} \right]\exp( ik_{bx}x + ik_{by}y),
\end{equation}
where $X_{0} = 8p$ is the beam width and $(k_{bx},k_{by})$ are the components 
of the incident Fourier components.  But first, let us study the two dimensional analogue of Figure \ref{fig2}. To this end, we assume $\alpha = 0.5$, $\beta = 0.2$ and the initial condition described by Eq. \eqref{eq25}.  The Fourier amplitudes evolution for the particular inputs centered at  
$(k_{x},k_{y}) = (\pm1,\pm1)$ is depicted in Figure \ref{fig10}, 
as the propagation distance $z$ increases. We clearly note the typical $\mathcal{PT}$ asymmetry also present in the one dimensional counterpart. 
In Figure \ref{fig11} we plot the same Fourier amplitudes as in 
Figure \ref{fig10} but with the lattice at the symmetry breaking point 
$\beta = \beta_{c}= 0.5$. On the right part of this figure, one can see 
that the lattice still supplies the field with optical energy. The left part of Figure \ref{fig11} however indicates that the initial mode located at $(k_{x},k_{y}) = (1,1)$ remains trapped in its initial configuration. 
Therefore, one may conclude that mode 
trapping and asymmetric mode power transfer is also exhibited by two dimensional complex lattices.

\section{Conclusions}

To conclude, based on the two-waves model, a theoretical study on the propagation of a wide beam through a transversal periodic photonic lattice described by a $\mathcal{PT}$ symmetric refractive index is developed, focusing on the behavior of power oscillations of a pair of resonant Bragg modes. As expected, within the paraxial wave approximation, a symmetry breaking phase transition is obtained by varying a critical parameter $\beta$ that measures the depart of the system from its Hermitian character. We have shown that small changes in this order parameter affect dramatically the dynamics and there are three regimes determined by $\beta$: for $\beta< \beta_c$ the beam power oscillates between the two modes; for $\beta>\beta_c$ it increases exponentially in a particular mode; while for $\beta = \beta_c$ depending on the initial condition, it remains constant or it grows as a quadratic function. Thus, we obtain power oscillations in the symmetric phase, and biased mode trapping at the critical point. It should be noted here that, the same behavior has been reported previously \cite{kottos} as a universal behavior typical of phase transitions, due to the fact that microscopic details are not important. Therefore, the results found here should apply to a wide class of $\mathcal{PT}$ systems.  To investigate the validity of the approximation we have verified our results and their limitations by simulating the propagation of finite width Gaussian pulses. Furthermore, extending the model to a two-dimensional wave equation, we have also observed power oscillations and power trapping. Although the results obtained here are based on a simple model, we have shown that they are exceptionally reliable. Due to its relative simplicity we hope that the present model might offer insights into the study of systems with generalized $\mathcal{PT}$ symmetries and an easy way for testing $\mathcal{PT}$ concepts. Considering the enormous potential of experimental researchers to produce artificial materials, one might conjecture that $\mathcal{PT}$-symmetric artificial systems should unravel a new generation of optical materials and consequently, of optical devices.

\section*{Acknowledgements}
The authors would like to acknowledge the Brazilian Agencies CNPq (457520/2014-0) and CAPES for financial support.


\end{document}